\documentclass[aps,prb,twocolumn,superscriptaddress,groupedaddress,floatfix,showpacs]{revtex4}
\usepackage{amsmath}
\usepackage{amsfonts}
\usepackage{amssymb}
\usepackage{graphicx}
\usepackage{bm}

\begin{document}

\title{Effect of Uniaxial Strain on Ferromagnetic Instability and Formation of Localized Magnetic States on Adatoms in Graphene}
\author{Anand Sharma} 
\email{anand.sharma@uvm.edu}

\author{Valeri N. Kotov}
\affiliation{Department of Physics, University of Vermont, 82 University Place, Burlington, Vermont 05405, USA}

\author{Antonio H. Castro Neto}
\altaffiliation{On leave from Department of Physics, Boston University, 590 Commonwealth Avenue, Boston, Massachusetts 02215, USA}
\affiliation{Graphene Research Centre and Department of Physics, National University of Singapore, 2 Science Drive 3, Singapore 117542}

\begin{abstract}
We investigate the effect of an applied uniaxial strain on the ferromagnetic instability due to long- range Coulomb interaction between 
Dirac fermions in graphene. In case of undeformed graphene the ferromagnetic exchange instability occurs at sufficiently strong 
interaction within the Hartree- Fock approximation. In this work we show that using the same theoretical framework but with an additional 
applied uniaxial strain, the transition can occur for much weaker interaction, within the range in suspended graphene. We also study the consequence of strain on the formation of 
localized magnetic states on adatoms in graphene. We systematically analyze the interplay between the anisotropic (strain- induced) 
nature of the Dirac fermions in graphene, on- site Hubbard interaction at the impurity and the hybridization between the graphene and 
impurity electrons. The polarization of the electrons in the localized orbital is numerically calculated within the mean- field self- 
consistent scheme. We obtain complete phase diagram containing non- magnetic as well as magnetic regions and our results can find 
prospective application in the field of carbon- based spintronics.
\end{abstract}

\date{\today}

\maketitle

\section{\label{sec:int} Introduction}

\indent Magnetism in organic (carbon- based) materials\cite{veciana,makarova} is not only an interesting physical property for studying 
from the fundamental point of view but it is also crucial in realizing carbon- based room- temperature spintronics\cite{wolf}, where one 
utilizes the charge as well as spin degrees of freedom of electrons. Carbon is long lasting as well as abundant, and  devices made out 
of carbon are expected to be advantageous in many ways and inexpensive. Moreover, there are many advantages of using organic instead of inorganic 
substances to make spintronics devices. The tunability of electronic properties, mechanical flexibility and weak spin scattering mechanisms 
in carbon- based materials could improve the reliability of this emergent technology.\\ 
\indent Carbon exists in many allotropic forms. Its two dimensional (2D) form, graphene, consists of carbon atoms arranged on a honeycomb 
lattice and in 2004, it was isolated for the first time from graphite in a controlled manner\cite{novoselov}. This naturally occurring 
crystalline pure material provides an avenue of many intriguing quantum phenomena due to its low dimensional structure. And its exceptional 
mechanical\cite{lee}, electronic\cite{castroneto1}, transport\cite{peres1} and optical\cite{nair} properties has made it a promising material 
for various practical applications. Recently electrical spin current injection and detection in graphene was established up to room 
temperature\cite{tombros}. And owing to the absence of nuclear magnetic moments in carbon and negligible spin- orbit 
interaction\cite{huertas-hernando}, graphene is a also good candidate for spintronics. \\
\indent But the lack of finite electronic band gap\cite{bostwick} has been a hindrance to use graphene for technological purposes. Since 
it has outstanding mechanical properties and is capable of sustaining huge atomic distortions\cite{lee}, there have been proposals to use 
uniaxial strain as a tool in order to open a band- gap and tailor the electronic properties of graphene. Pereira \textit{et al.}\cite{pereira1} 
analyzed such concepts within the standard tight- binding (TB) approximation and effect of strain was also studied experimentally using 
Raman spectroscopy\cite{ni} or by using metal cluster super-lattices and patterned modifications\cite{park}. Although uniaxial strain can not 
generate the required bulk band- gap but there are other important consequences of it on graphene besides adjusting the electronic 
properties\cite{goerbig}. For instance, deformations due to strain can alter its optical properties\cite{pellegrino1}, modulate its specific 
heat\cite{xia}, affect the plasmon excitations\cite{pellegrino2} and exhibit rich interplay with electron- electron interactions\cite{sharma}.\\
\indent For technological applications, it is also desirable to understand and control its magnetic properties besides tunable electronic 
and optical properties. In order to investigate the intrinsic magnetism in graphene, Peres \textit{et al.}\cite{peres2} 
analyzed the exchange instabilities induced by the long- range Coulomb interaction between the Dirac fermions within the Hartree- Fock (HF) 
approximation. Their results showed that in undoped graphene, the itinerant electrons exhibit a ferromagnetic transition when the coupling constant is 
sufficiently large. However according to recent experiments\cite{sepioni}, no evidence for intrinsic magnetism was found at any temperature 
down to 2 K. This might be a consequence of weak effective interaction\cite{reed} between the Dirac fermions and is a subject of on- going 
research.\\
\indent Apart from ferromagnetic exchange instabilities due to long- range interactions, there are numerous studies on the induced localized 
magnetism due to on- site Hubbard- type interactions in finite sheets (nanoribbons\cite{yazyev}) and with dopants (vacancies and external 
adatoms\cite{castroneto2}) in graphene. Uchoa \textit{et al.}\cite{uchoa1} examined the possibility of local moment formation on adatom 
(impurity ion) in graphene. Their analysis was based on Anderson's single impurity model\cite{anderson} in which the constant density of 
states of the impurity hybridizes with the sea of conduction Dirac electrons in graphene. They concluded that under certain conditions 
(physical parameters of the model) it is feasible to form a localized magnetic moment on the impurity. They also showed that the local 
magnetic moment can be tuned by applying a potential through an electric field via back gate, emphasizing the importance of electrically- 
controlled magnetic properties of an impurity in graphene. \\
\indent The purpose of this work is to explore the effect of an uniaxial strain not only on the itinerant magnetism i.e., ferromagnetic 
exchange instability due to long- range Coulomb interaction but also on the localized magnetism i.e., formation of localized magnetic 
moment due to an impurity in graphene. There are studies which aimed towards understanding magnetism in strained graphene but they are based 
on finite size nanoribbons\cite{viana-gomes}, topological line defects\cite{kou}, transition- metal atoms adsorbed on graphene based on density 
functional theory calculations\cite{huang} or modulation of magnetic (RKKY-type) interactions between localized moments in graphene\cite{power}.\\
\indent The rest of the paper is structured as follows. In section~\ref{sec:aed} we present the tight- binding Hamiltonian in the presence 
of an uniaxial strain and derive the anisotropic electronic dispersion of graphene. In section~\ref{sec:fei} we examine the possibility 
of ferromagnetic exchange instabilities due to long- range Coulomb interaction between anisotropic Dirac fermions under applied strain 
within the HF approximation. We study the change in the strength of effective critical interaction or coupling constant which causes 
the magnetic transition. In section~\ref{sec:lms} we formulate the conditions for the existence of localized magnetic moments on an 
adatom in a deformed graphene. We present phase diagrams for different values of model parameters. In the last section~\ref{sec:con} 
we summarize and conclude our findings.\\

\section{\label{sec:aed} Anisotropic Electronic Bandstructure}

\begin{figure}
\centering
\includegraphics[height=5.0cm,width=7.5cm]{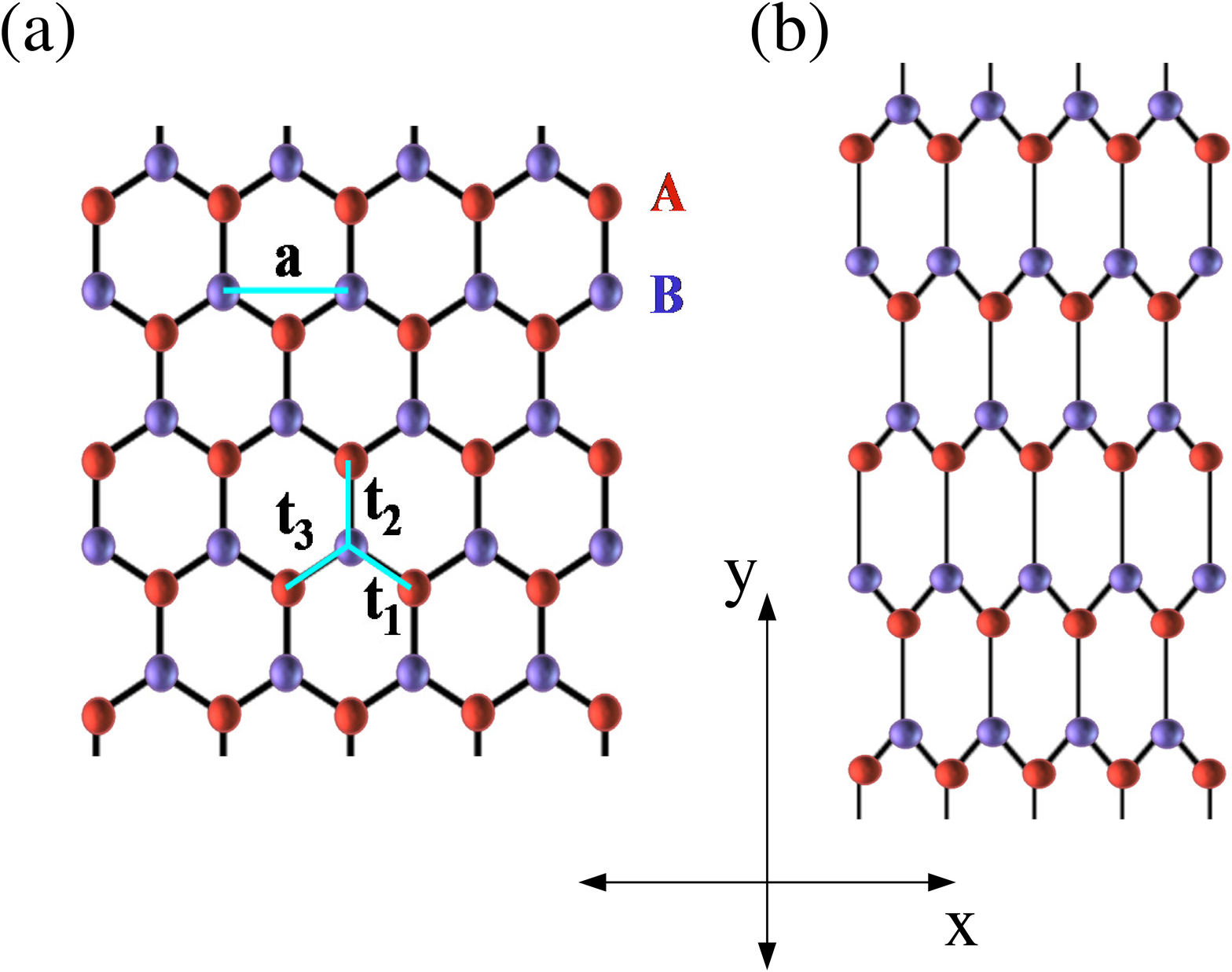}
\caption{\label{fig:1} (Color online) Left : Undeformed graphene honeycomb lattice structure showing two sub- lattices A and B with t$_{1}$, t$_{2}$ 
and t$_{3}$ as the nearest neighbor hopping parameters. Right : Uniaxially strained along the armchair, $y-$ direction.}
\end{figure}

\indent We begin by deriving the expression for the electronic dispersion of deformed graphene. In the undeformed case, 
P. R. Wallace\cite{wallace} was the first to study its band structure, as early as 1947, in terms of tight- binding model. In graphene the 
carbon atoms are arranged in a two dimensional (2D) honeycomb lattice, see Fig.~\ref{fig:1}(a), with two atoms (A and B) in the unit cell. 
The Hamiltonian as obtained from tight- binding model reads
\begin{equation}{\label{eq:1}}
H_{0} = \sum_{\textbf{R},\gamma_{i},\sigma} t(\gamma_i) \textrm{a}^{\dagger}_{\textbf{R}\sigma} \textrm{b}_{\textbf{R}+\gamma_{i},\sigma} + H.c 
\end{equation}
where $t(\gamma_i) = t_i$ are the nearest neighbor hopping parameters as shown in Fig.~\ref{fig:1}(a). In the isotropic case they are 
all equalivalent ; $t_0 = t_i \approx 2.7 eV$. Here a$^{\dagger}_{\textbf{R}\sigma}$ (b$_{\textbf{R}+\gamma_{i},\sigma}$) is creation 
(annihilation) operator of an electron at sub- lattice A (B) with spin $\sigma = \uparrow , \downarrow$ at position 
$\textbf{R} \hspace{0.1cm} (\textbf{R}+\gamma_{i})$ and defined as 
a$^{\dagger}_{\textbf{R}\sigma} = \frac{1}{N} \sum_{\textbf{k}} e^{-i\textbf{k}\cdot\textbf{R}} \textrm{a}^{\dagger}_{\textbf{k}\sigma}$ 
(b$_{\textbf{R}+\gamma_{i}\sigma} = \frac{1}{N} \sum_{\textbf{k}} e^{i\textbf{k}\cdot(\textbf{R}+\gamma_{i})} \textrm{b}_{\textbf{k}\sigma}$). 
The bond vectors from atom A (or B) to its nearest neighbor B (or A) are denoted by $\gamma_{i}$ with $\gamma_{1} = a(\frac{\sqrt{3}}{2}, \frac{-1}{2})$, 
$\gamma_{2} = a(0,1)$ and $\gamma_{3} = a(\frac{-\sqrt{3}}{2}, \frac{-1}{2})$. The bond length or the distance between two carbon atoms is 
$a (\approx$0.142 nm) and a = $\sqrt{3} a$ as shown in Fig.~\ref{fig:1}(a). We consider uniform bond deformations and neglect the bond 
bending effects. \\

\indent In the momentum space the Hamiltonian in Eq.~(\ref{eq:1}) can be written as
\begin{equation}{\label{eq:2}}
H_{0} = -t_{2} \sum_{\textbf{k}\sigma} \phi(\textbf{k}) a^{\dagger}_{\textbf{k}\sigma} b_{\textbf{k}\sigma} + H.c
\end{equation}
where $\phi(\textbf{k}) = e^{i\textbf{k} \cdot \textbf{$\gamma$}_{2}} [1 + 2\eta e^{-ik_{y}a_{y}} \textrm{cos}(k_{x}a_{x})]$, 
$\eta = \frac{t_1}{t_2} = \frac{t_3}{t_2} , a_{x} = \frac{\textrm{a}}{2}$ 
and $a_{y} = \frac{\sqrt{3}}{2}$a. We obtain the low- energy effective Hamiltonian by expanding the wave vector, $\textbf{k}$, near one of the 
six Dirac points. In fact there are only two, out of six Dirac points, which are inequivalent. Therefore on expanding around either of the 
two inequivalent Dirac points, $(\pm\frac{4\pi}{3\sqrt{3}}, 0)$ we get
\begin{equation}{\label{eq:3}}
\hat{H}_0 = v_{x}k_{x}\hat{\tau}_{x} + v_{y}k_{y}\hat{\tau}_{y}
\end{equation}
where $\hat{\tau}_{x}$, $\hat{\tau}_{y}$ are the  usual $(2\times2)$ Pauli matrices. Here $v_{x} = t_{2} a_{x} \sqrt{4 \eta^2 -1}$, 
$v_{y} = t_{2} a_{y}$ are the velocities along the $x-$ and $y-$ directions respectively. And on diagonalizing the Hamiltonian in 
Eq.~(\ref{eq:3}), we get the required anisotropic electronic dispersion (elliptical double cones)\cite{goerbig} 
\begin{equation}{\label{eq:4}}
\varepsilon({\bf k}) = \pm\sqrt{v_x^2k_x^2 + v_y^2k_y^2} 
\end{equation}
where $\pm$ denotes the electron- and hole- like bands. We define momentum cut- offs along $x-$ and $y-$ directions in such a way that the number 
of states in the Brillouin zone are conserved while going from isotropic to the anisotropic case i.e., the area of the unit cell remains constant. 
Thus $k^x_c = \frac{D_a}{v_x} = \sqrt{\frac{v_y}{v_x}}k_c$ and $k^y_c = \frac{D_a}{v_y} = \sqrt{\frac{v_x}{v_y}} k_c$ where $D_a$ is the anisotropic 
energy cut- off and $k_c$ being the isotropic momentum cut- off. And the area, $A$, of the unit cell is defined so that 
$\frac{(2\pi)^2}{\pi A} = \frac{D^{2}_{a}}{v_x v_y} = k^x_c k^y_c = k^{2}_{c}$. \\
\indent In the following sections, we inspect the conditions required to observe an exchange instability due to long- range Coulomb interaction 
(section~\ref{sec:fei}) and the possibility of forming a localized magnetic moment on an impurity (section~\ref{sec:lms}) in deformed graphene.\\

\section{\label{sec:fei} Ferromagnetic Exchange Instability }

\indent In 1929, Bloch proposed the occurence of ferromagnetic transition driven by electronic density in a three- dimensional (3D) 
interacting electron gas using Hartree- Fock mean- field theory\cite{bloch}. The result was that at high density the electron system 
would have paramagnetic ground state in order to optimize the cost in kinetic energy due to large number of electrons while at low density 
the system should spontaneously spin- polarize into a ferromagnetic ground state so as to compromise for the exchange energy arising from the 
Pauli principle and Coulomb interaction. \\
\indent For an electron gas in a positive background, it is easy to calculate\cite{giuliani} the total HF energy per particle (which is a 
sum of the non- interacting kinetic energy and the Fock exchange energy due to unscreened Coulomb interaction at zero temperature) and 
see that it leads to first- order ferromagnetic transition, so- called Bloch transition. This happens when the dimensionless parameter, 
$r_{s}$ which is defined as the ratio of potential to the kinetic energy and which is inversely proportional to the electronic density, 
becomes greater than a critical value. In such a situation, the ferromagnetic (fully spin- polarized) state is lower in energy than the 
paramagnetic (unpolarized) state. \\
\indent In 2005, Peres \textit{et al.}\cite{peres2} investigated the possibility of such a transition in pure (half- filled undoped) 
graphene within the Hartree- Fock approximation. They concluded that for coupling constant, $\alpha$, greater than certain critical 
value, $\alpha_{c} \approx$ 5.3, the ground state of the system becomes maximally spin- polarized. Since the behavior of electrons in 
graphene is quite different than the usual two- dimensional electron gas, one might argue that it is inadequate to calculate the 
ferromagnetic exchange instability based on the Hartree- Fock approximation and one has to include the correlation effects i.e., going 
beyond the Hartree- Fock approximation. Such an estimate of the correlation energy was made by Dharma-wardana\cite{dharma-wardana}, 
The author found out that the inclusion of correlations suppressed the exchange- driven spin- polarized phase. But undoped 
graphene has vanishing density of states near the Fermi energy (Dirac point) and the long- range Coulomb interaction remains  unscreened. 
Moreover unlike usual electron gas system, the coupling constant in graphene is independent of its electronic density. So it is an elusive 
task to quantitatively establish the correlation effects in graphene. Following Ref.~\onlinecite{peres2}, we examine the effect of 
uniaxial strain on the ferromagnetic exchange instabilities in graphene within Hartree- Fock approximation. Even though it must be kept in mind that at Hartree-Fock level the tendency toward itinerant magnetism is possibly too strong, as is the case in the conventional
electron gas \cite{giuliani}, this level of approximation allows for complete analysis of the effects due to strain (Dirac fermion anisotropy).\\
\indent 
We consider the case of charge neutral graphene, i.e. zero chemical potential. It is expected that, as in unstrained graphene \cite{peres2}, 
 the ferromagnetic instability is the strongest in this case. 
The ground state energy per particle, per valley, of a spin polarized system is expressed in terms of the kinetic, exchange and correlation 
energies. The kinetic energy, from Eq.~(\ref{eq:4}), is given by
\begin{eqnarray}{\label{eq:5}}
K & = & g_s \hbar \sum_{k} \sqrt{v_x^2k_x^2 + v_y^2k_y^2} \nonumber \\
& = & g_s \hbar A \iint \sqrt{v_x^2k_x^2 + v_y^2k_y^2} \frac{d^{2}k}{(2\pi)^{2}}
\end{eqnarray}
where $g_s$=2 is spin degeneracy factor. 
 It is straight- forward to derive the kinetic energy of the unpolarized (paramagnetic) state, $K_p$, as a 
function of isotropic momentum cut- off, $k_c$ ;
\begin{equation}{\label{eq:6}}
K_p = - \frac{g_s \hbar A}{6\pi} \sqrt{v_x v_y} k^3_c\\
\end{equation}
and also the kinetic energy of the polarized (ferromagnetic) state, $K_f$, for either of the spin as a function of $k_c$ and the 
Fermi wavevector, $k_F$ ;
\begin{equation}{\label{eq:7}}
K_f = - \frac{\hbar A}{6\pi} \sqrt{v_x v_y} (k^3_c-k^3_F)\\
\end{equation}
Here the Fermi wavevector $k_F$ is related to the magnetization $M$ via: 
$M = \frac{n_\uparrow - n_\downarrow}{n_\uparrow + n_\downarrow} = (\frac{k_F}{k_c})^2$ with $n_\sigma$ being 
the occupation number for spin $\sigma$.
The kinetic energy difference, $\Delta K$ , between the ferromagnetic and paramagnetic 
phases is given as
\begin{eqnarray}{\label{eq:8}}
\Delta K & = & g_s K_f - K_p \nonumber \\
& = & \frac{g_s \hbar A}{6\pi} \sqrt{v_x v_y} k^3_F
\end{eqnarray}
where the spin degeneracy factor takes into account the individual contributions of either of the spins for the spin- polarized case.\\
\indent We now evaluate the interaction energy which consists of both exchange and the correlation terms. But as mentioned earlier, we 
shall calculate only the exchange energy term and neglect the correlation energy. In graphene, the electrons interact via long- range 
Coulomb interaction which is defined as $V_{\textbf{p}} = \frac{2 \pi e^2}{\kappa |\textbf{p}|}$ where $\kappa$ is the appropriate dielectric 
constant. The exchange energy is determined by electron- hole excitations and within the Hartree- Fock approximation is given by
\begin{equation}{\label{eq:9}}
E_{ex} = - \frac{Tr}{2} \sum_{\textbf{k},\textbf{p},\sigma} \int \frac{d\omega}{2\pi} \int \frac{d\epsilon}{2\pi} V_{\textbf{k-p}} \hspace{0.2cm} i 
\hat{G}^0_{\sigma}(\textbf{k},\omega) \hspace{0.2cm} i \hat{G}^0_{\sigma}(\textbf{p},\epsilon) \\
\end{equation}
where $\hat{G}^0_{\sigma}(\textbf{k},\omega)$ is the non- interacting Green's function for spin $\sigma$ which is defined as
\begin{equation}{\label{eq:10}}
\hat{G}^0_{\sigma}(\textbf{k},\omega) = \frac{\omega \hat{\tau}_0 + v_xk_x\hat{\tau}_x + v_yk_y\hat{\tau}_y}{\omega^2 - (v_x^2k_x^2 + v_y^2k_y^2)}
\end{equation}
with $\hat{\tau}_0$ being the $(2\times2)$ identity matrix and $\hat{\tau}_x$, $\hat{\tau}_y$ are the Pauli matrices. The momentum 
summations in Eq.~(\ref{eq:9}) are reduced to integrals with limits of the integrations as $k_c$ or $k_F$ depending on whether the 
exchange energy is calculated for paramagnetic or ferromagnetic phase. After making use of scaling transformations we evaluate the 
exchange energy difference, $\Delta E_{ex}$, between the ferromagnetic and paramagnetic phases as
\begin{equation}{\label{eq:11}}
\Delta E_{ex} = -\frac{\hbar A \sqrt{v_x v_y} k^3_F \alpha_x}{16 \pi^3} \left(I_1 - \frac{2I_2}{\sqrt{M}}\right)  \\
\end{equation}
where $\alpha_x = \frac{e^2}{\kappa v_x}$ is the strength of the coupling constant which is also given as the ratio of potential 
to the kinetic energy along $x-$ direction with the directional dependence coming because of anisotropy, and
%
\begin{equation}{\label{eq:12}}
I_1 = \int^1_0 \int^1_0 \int^{2\pi}_0 \int^{2\pi}_0 \frac{r t (1 + \textrm{cos}(\theta-\phi))dr dt d\theta d\phi}{f(r,t,\theta,\phi)} \\
\end{equation}
and
\begin{equation}{\label{eq:13}}
I_2 = \int^1_0 \int^1_0 \int^{2\pi}_0 \int^{2\pi}_0 \frac{r t \textrm{cos}(\theta-\phi)dr dt d\theta d\phi}{g(r,t,\theta,\phi)} \\
\end{equation}
are the integrals which depend on the functions 
\begin{equation}
f(r,t,\theta,\phi) = \sqrt{\left(\frac{v_y}{v_x}(r\textrm{cos}\theta - t\textrm{cos}\phi)\right)^2 + (r\textrm{sin}\theta - t\textrm{sin}\phi)^2} \nonumber 
\end{equation}
and 
\begin{eqnarray*}
\lefteqn{g(r,t,\theta,\phi) =} \\
&&\sqrt{\left(\frac{v_y}{v_x}(r\sqrt{M}\textrm{cos}\theta - t\textrm{cos}\phi)\right)^2 + (r\sqrt{M}\textrm{sin}\theta - t\textrm{sin}\phi)^2} \\
\end{eqnarray*}
\indent In what follows we shall define the anisotropy parameter due to applied uniaxial strain. We define it to be proportional to 
the ratio of the two velocities, $v_{y}$ and $v_{x}$ i.e., let $\frac{v_{y}}{v_{x}} \equiv 1-\delta$. We are interested in the 
range $0\leq \delta \leq 1$, i.e. $1 \geq  v_y/v_x \geq 0$, which according to Fig.~\ref{fig:1}(b) suggests that the tension is along 
the armchair, $y-$ direction with $v_y< v_x$. The case $\delta=0$ corresponds to the isotropic and $\delta=1$ is the limit of decoupled 
chains of carbon atoms. In the similar manner we can describe strain along zig- zag, $x-$ direction, by using the same parameter range 
and accordingly relabeling the axes. \\
\indent The total energy gain between the ferromagnetic and paramagnetic phase is given by the sum of the kinetic, Eq.~(\ref{eq:8}), and 
exchange energy differences, Eq.~(\ref{eq:11});
\begin{eqnarray}{\label{eq:14}}
\Delta E & = & \Delta K + \Delta E_{ex} \nonumber \\
& = & \hbar A \sqrt{v_x v_y} k^3_F \left[ \frac{1}{3\pi} - \frac{\alpha_x}{16\pi^3}\left(I_1 - \frac{2I_2}{\sqrt{M}}\right)\right]
\end{eqnarray}
where the integrals are functions of anisotropy and magnetization. \\
\indent We find numerically that the exchange energy exceeds the kinetic energy for large magnetization $M=1$ (i.e.
 $k_F  = k_c$), which leads to the interaction-driven 
ferromagnetic transition.  The case of partial spin polarization $M < 1$ is always energetically less favorable.
 The large magnetization is  characteristic feature of the Hartree-Fock approximation
\cite{peres2,giuliani}. 
 Such a scenario occurs when the coupling constant strength, $\alpha_x$, is greater than certain critical 
value, $\alpha^{c}_{x}$ ;
\begin{equation}{\label{eq:15}}
\alpha_{x} \geq \alpha^{c}_{x} = \frac{16 \pi^2}{\left(3I_1 - \frac{6I_2}{\sqrt{M}}\right)}
\end{equation}
Thus when the condition in Eq.~(\ref{eq:15}) is satisfied for a given anisotropy, 
the system exhibits a first-order transition (Bloch transition) and the spin degeneracy is lifted, i.e.
for $\alpha_x < \alpha^{c}_{x}$ the system is unpolarized (paramagnetic) and when $\alpha_x > \alpha^{c}_{x}$ it is fully spin- 
polarized (ferromagnetic).\\
\begin{figure}
\centering
\includegraphics[height=5.0cm,width=7.5cm]{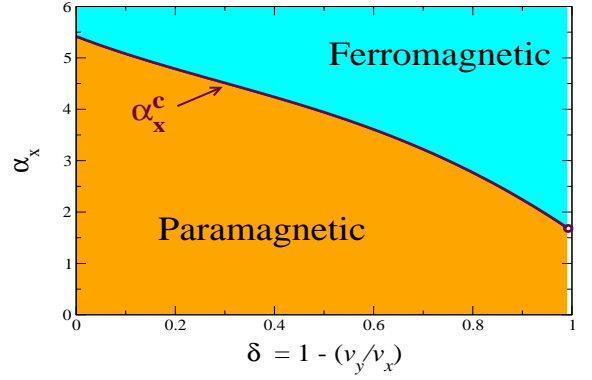}
\caption{\label{fig:2} (Color online) Boundary separating the paramagnetic and ferromagnetic regions indicating the onset of an 
exchange instability in deformed graphene. The value of the critical coupling constant, $\alpha^{c}_{x}$, decreases as a function 
of applied uniaxial strain.}
\end{figure}
\indent We solve for the critical values of coupling constant for $M = 1$ and plot it as a function of anisotropy, 
$\delta = 1 -\frac{v_y}{v_x}$, as shown in Fig.~\ref{fig:2}. The phase boundary, where the Bloch instability occurs, separates the 
unpolarized from the fully spin- polarized region. Our main result is that with an applied uniaxial strain (increasing anisotropy) 
the value of the critical coupling constant, $\alpha^{c}_{x}$, decreases. In the limit of extreme strain, $\delta = 1$, the two- 
dimensional graphene decouples into chains of carbon atoms. It is found that the value of critical coupling constant is finite and 
is depicted as a circle in Fig.~\ref{fig:2}. The reason $\alpha^{c}_{x}$ decreases with increasing anisotropy
is due to the fact that electronic instabilities are generally enhanced in lower dimensions. However we should still
 keep in mind that even for large anisotropy  the critical interaction strength remains relatively large $\alpha^{c}_{x} \lesssim 2$ and thus correlation energy effects could be substantial. Nevertheless our result demonstrates that
 itinerant ferromagnetism is strongly favored in strained graphene - an exciting possibility not yet tested 
 experimentally. \\

\section{\label{sec:lms} Localized Magnetic States on Adatom}

\begin{figure}
\centering
\includegraphics[height=4.5cm,width=8.5cm]{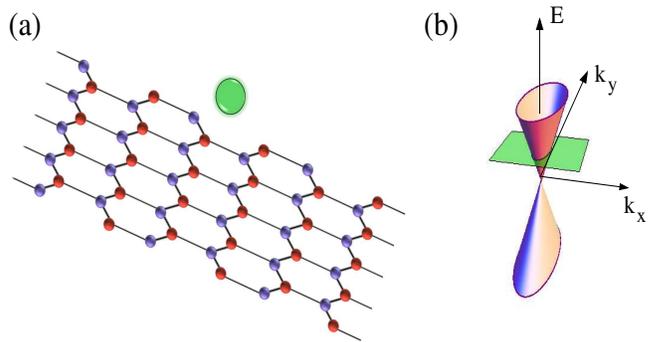}
\caption{\label{fig:3} (Color online) Schematic of (a) an uniaxially strained graphene with an adatom (in green) sitting on top of 
one of the two sub- lattices (red or blue) and (b) the corresponding anisotropic dispersion of the Dirac fermions near one of the 
Dirac points along with flat dispersionless level of the localized adatom which can be moved in the conduction (as shown in the figure) 
or in the valence band.}
\end{figure}

\indent In this section we investigate the conditions required to form localized magnetic moments on an adatom in strained graphene. 
Our work closely follows the analysis of Uchoa \textit{et al.}\cite{uchoa1} in terms of the description of the model using the single 
impurity Anderson Hamiltonian. Although their investigation was extended for the situation with magnetic adatom\cite{cornaglia} and 
for an impurity on bilayer graphene\cite{killi}, the case of deformed monolayer graphene with a single non- magnetic adatom 
remains unexplored.\\
\indent We consider an uniaxially strained graphene sheet with a single adatom (impurity ion shown in green) sitting on top of 
one of the two inequivalent atoms (A or B shown in red or blue respectively) as shown in Fig.~\ref{fig:3}(a). The anisotropic 
dispersion (elliptical cone) for one of the valley degrees of freedom along with flat dispersionless level of the localized 
impurity in the conduction band is presented in Fig.~\ref{fig:3}(b). The model Hamiltonian for such a system in momentum space 
is expressed as
\begin{equation}{\label{eq:16}}
H = H_{0} + H_{f} + H_{V}
\end{equation} 
where the first term , $H_{0}$, is the anisotropic dispersion of the Dirac electrons in graphene as given in Eq.~(\ref{eq:2}). 
The second term describes the featureless localized orbital of an impurity ion (constant density of the states at energy $E_{0}$) 
along with the Coulomb repulsion energy which takes into account the double occupancy of the localized energy level,
\begin{equation}{\label{eq:17}}
H_{f} =  \sum_{\sigma} (E_{0} + U n^{f}_{-\sigma}) f^{\dagger}_{\sigma}f_{\sigma} 
\end{equation} 
where $f^{\dagger}_{\sigma}$ $(f_{\sigma})$ creates (annihilates) an electron with spin $\sigma$ at the impurity site and we 
have used the mean- field decoupling of the local interaction\cite{uchoa1}, $U$. The spin- dependent occupation number of the 
electrons at the impurity level, $n^{f}_{\sigma}$, depends on its density of states and will be defined in the following text. 
The localized level is populated by the Dirac electrons due to hybridization which is given by the last term in Eq.~(\ref{eq:16}),
\begin{equation}{\label{eq:18}}
H_{V} = V\sum_{\textbf{p}\sigma} (f^{\dagger}_{\sigma}a_{\textbf{p}\sigma} + a^{\dagger}_{\textbf{p}\sigma} f_{\sigma}) 
\end{equation} 
where $V$ is the strength of hybridization and we assume that the impurity sits on top of atom A.\\
\indent We use the Green's function method to solve the Hamiltonian in Eq.~(\ref{eq:16}). Since we are interested in the condition 
of formation of localized magnetic moment on adatom, we consider the full interacting Green's function of the electrons in the 
localized level which is defined as $G_{f\sigma}(\omega) = [\omega - \epsilon_{\sigma} - \Sigma_{f} (\omega) + i0^{+}]^{-1}$ where 
$\epsilon_{\sigma} = E_{0} + U n^{f}_{-\sigma}$ is the energy of the localized electrons with spin $\sigma$ in the presence of 
interaction energy $U$ and the self-energy has the form,  
\begin{eqnarray}{\label{eq:19}}
\Sigma_{f}(\omega) & = & V^{2} \sum_{\textbf{p}} \frac{\omega}{\omega^{2} -[(v^{x}p_{x})^2 +(v^{y}p_{y})^2]} \nonumber \\
& = & \frac{V^2}{D^2_a} \left[\omega\textrm{log}\left(\frac{\omega^2}{D^2_a - \omega^2}\right) - i \pi |\omega| \Theta(D_a - |\omega|) \right] \nonumber \\
\end{eqnarray}
where $D_a$ is the anisotropic energy cut- off as defined in section~\ref{sec:aed} and $\Theta$ being the Heaviside step function. 
With the knowledge of the full Green's function we can obtain the density of states of the impurity electrons in the localized state,
$\rho_{f\sigma}(\omega) = -\frac{Im}{\pi} G_{f\sigma}(\omega)$. \\
\begin{figure*}
\centering
\includegraphics[height=12cm,width=18.0cm]{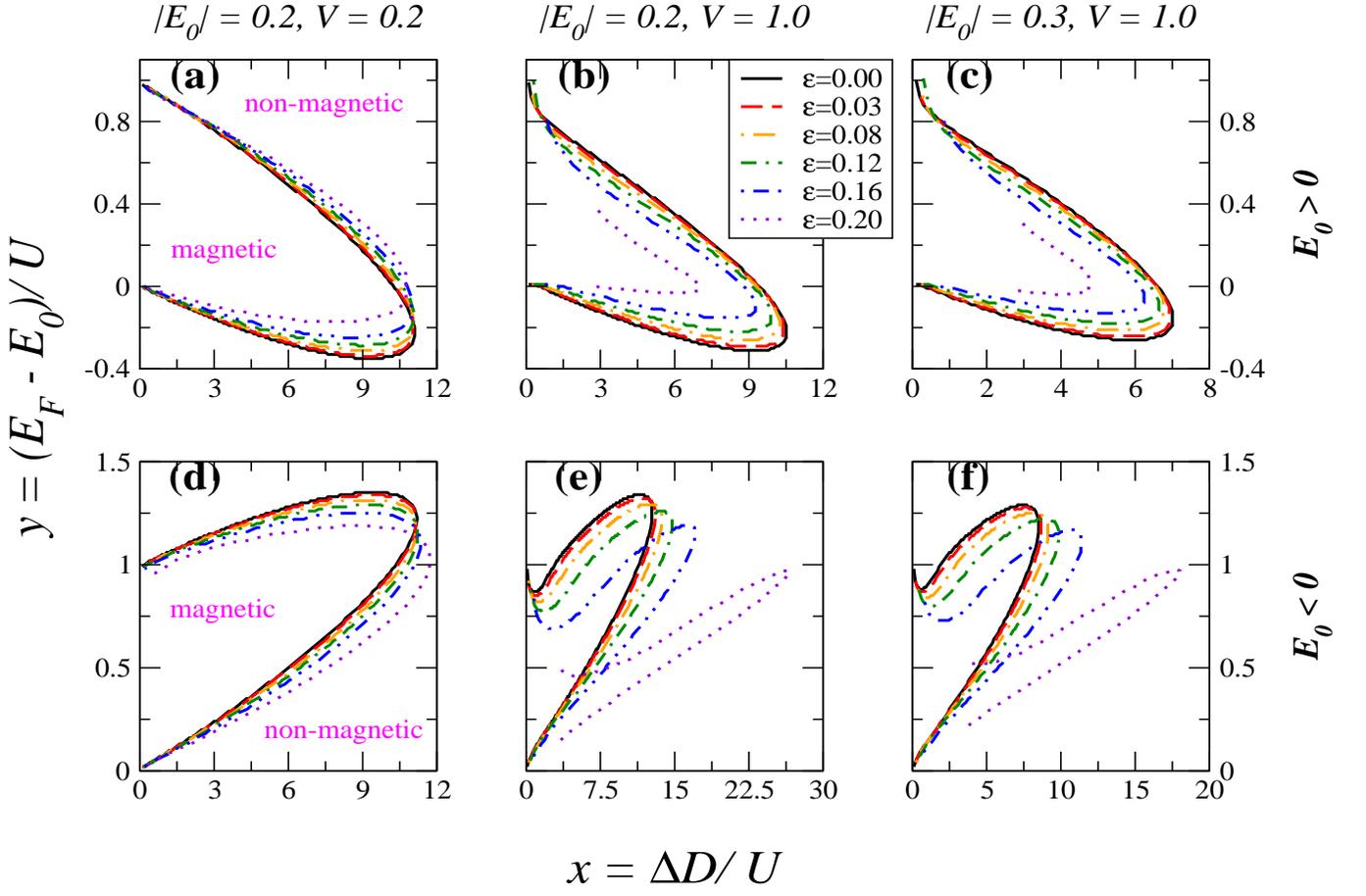}
\caption{\label{fig:4} (Color online) Phase diagram showing the magnetic regions signaling the onset of formation of localized 
magnetic states on the adatom (impurity) in graphene for various values of applied uniaxial strain. The $x-$ and the $y-$ axis 
of the phase diagram are scaled model parameters. Each phase plot is for a fixed value of localized impurity energy level ($E_{0}$) 
and the strength of bare hybridization ($V$) in units of eV. The plots in the upper (lower) panels are for positive (negative) 
values of $E_{0}$.}
\end{figure*}
\indent The condition to form a magnetic moment on the impurity is determined by calculating the spin- dependent occupation number 
which is given by
\begin{eqnarray}{\label{eq:20}}
n^{f}_{\sigma} & = & \int^{E_F}_{-D_{a}} \rho_{f\sigma} (\omega) d\omega \nonumber \\
& = & \frac{\Delta}{\pi} \int^{\frac{E_F}{D_{a}}}_{-1} \frac{|x|dx}{\left[ \left(x \left[ 1 - \frac{\Delta}{\pi} \textrm{log} (\frac{x^{2}}{1 - x^{2}}) \right] - \tilde{\epsilon_{\sigma}} \right)^{2} + \Delta^2 x^2 \right]} \nonumber \\
\end{eqnarray}
where $E_F$ is the Fermi energy which is tunable parameter by applying gate voltage, $\Delta = \frac{\pi V^{2}}{D^{2}_{a}}$ is 
the effective hybridization and $\tilde{\epsilon_{\sigma}} = \frac{E_{0} + U n^{f}_{-\sigma}}{D_{a}}$ is the energy of the localized 
electrons which is scaled with respect to the anisotropic cut- off. It is evident from Eq.~(\ref{eq:20}) that the determination of 
spin- dependent occupation number requires the self- consistent calculation of the density of states, $\rho_{f\sigma}(\omega)$, 
which incorporates the broadening of the impurity level due to hybridization with the sea of Dirac electrons in graphene. A 
localized moment is formed at the impurity whenever there is an imbalance in the spin- dependent occupation number at the level.\\
\indent We define anisotropy parameter due to strain through the anisotropic energy cut- off, $D_a$ which depends on velocities; 
$v_x$ and $v_y$. When we apply tension along the armchair ($y-$) direction as shown in Fig.~\ref{fig:1}(b) the bond vectors yields 
$ | \gamma_{1} | = | \gamma_{3} | = (1 + \frac{\varepsilon}{4} - \frac{3\varepsilon \nu}{4})a $ and $ | \gamma_{2} | = (1 + \varepsilon)a $ 
where we assume that the change in bond length is linear in strain\cite{pereira1} ($\varepsilon$) and $\nu = 0.165$ is the Poisson's 
ratio of graphite\cite{blakslee}. On taking the functional form of hopping parameters\cite{papaconstantopoulos} as 
$t_{i} = t_{0} e^{-\lambda a\left(\frac{| \gamma_{i} |}{a} - 1 \right)}$, we get the velocities as a function of applied strain 
which are given by $v_x = v_{F} (1 + \lambda a \varepsilon \nu) $ and $v_y = v_{F} (1 - \lambda a \varepsilon)$ where 
$v_F =\frac{3 t_{0} a}{2}$ is the isotropic Fermi veolcity. The anisotropic energy cut- off as a function of strain is given by, 
$D_a = \sqrt{v_x v_y} k_c \approx D (1 - c \varepsilon)$  with $D = v_F k_c \approx$ 7 eV being the isotropic graphene bandwidth 
and $c = \lambda a (1 - \nu)$ is a constant with $\lambda a \approx $ 3.37 as given in Ref.~\onlinecite{pereira1}. Notice that 
the anisotropy parameter is defined through the product of two velocities, $v_x$ and $v_y$, unlike in the previous section where 
it was a function of ratio of the two velocities.\\
\indent We regard $x = \frac{\Delta D}{U}$ and $y = \frac{(E_F - E_0)}{U}$ as scaled model parameters and solve Eq.~\ref{eq:20} 
for different values of impurity level ($E_0$), strength of bare hybridization ($V$) and local interaction ($U$) given in units 
of eV and for applied strain ($\varepsilon$). We obtain a phase diagram as shown in Fig.~\ref{fig:4} and observe clear boundary 
separating the magnetic and non- magnetic regions. Within our model the impurity level can be shifted either in the conduction 
band i.e., $E_0 > 0$ as shown in Fig.~\ref{fig:3}(b) or in the valence band i.e., $E_0 < 0$ and the corresponding phase plots 
are shown in the upper or lower panels of Fig.~\ref{fig:4} respectively. \\
\indent For the sake of completeness, we briefly summarize the main results of undeformed graphene ($\varepsilon = 0$) as obtained 
in Ref.~\onlinecite{uchoa1}. As seen in Fig.~\ref{fig:4}, the phase diagram is not symmetric for the cases where the localized 
energy level is either above or below the Dirac point and also around $y = 0.5$ which shows the particle- hole symmetry breaking 
due to existence of the impurity level. Moreover, when $E_0 > 0$ the magnetic boundary crosses the line $y = 0$ and the impurity 
magnetizes even when the energy level is above the Fermi energy. And similarly for the case $E_0 < 0$, the crossing occurs along 
$y = 1$ which suggests that even when the energy of the doubly occupied level is less than the Fermi energy the impurity gets 
magnetized. This happens due to the fact that the hybridization leads to large broadening of localized level density of states 
that crosses the Fermi energy. These features are very different as compared to the case of an impurity in a metal\cite{anderson}. 
In the following, we systematically study the effect of uniaxial strain on the the scaling of the magnetic phase boundary. \\
\indent We first consider the case when the localized level is in the conduction band ($E_0 > 0$) and show the results in 
Fig.~\ref{fig:4}(a) - (c). For a fixed strength of bare hybridization ($V = 0.2$) shown in Fig.~\ref{fig:4}(a), which is considered 
to be weak as compared to the graphene bandwidth, the magnetic region starts depleting as the applied strain increases. And this 
effect is more pronounced with stronger hybridization ($V = 1.0$) as shown in Fig.~\ref{fig:4}(b) and (c). The reason being with 
increasing strain the anisotropic cut- off energy ($D_a$) decreases and for given value of $V$, the effective hybridization ($\Delta$) 
increases. Thus the density of states at any given energy increases since it is proportional to the effective hybridization and this 
results in the electrons at the impurity level becoming delocalized. Moreover the delocalization increases upon enhancing the 
hybridization strength, $V$. \\
\indent It is also observed that for undeformed case, the size of the magnetic region gets enlarged as the impurity level 
($E_0$) approaches the zero- energy Dirac point for a given fixed hybridization. The formation of local moment is favored due 
to suppresion of density of states at low energies. If the value of bare hybridization is large then with strong deformations 
the magnetic region gets drastically reduced, as seen in Fig.~\ref{fig:4}(b) and (c), due to augmented density of states. 
In such a situation there is an emergence of a critical value of the scaled parameter ($x$) above which the impurity gets magnetized. This happens because the model breaks down for small $x$ and substantial anisotropy, since in this case
the Hubbard-shifted impurity level $E_0 +U$ is actually  outside the Dirac cone bandwidth.\\ 
\indent The depletion in the magnetic region with applied strain is also found for the case when the localized level is in the 
valence band ($E_0 < 0$) and the results are shown in Fig.~\ref{fig:4}(d) - (f). Apart from similar observations as in the case 
of $E_0 > 0$ , it is also seen that for large hybridization and small anisotropy, there's an up turn close to point 
$y = 1$ and $x \approx 0$ i.e., the magnetic phase boundary approaches that point from below. This suggests that in such a 
scenario the system behaves more like an impurity in a metal, i.e. the usual  single impurity Anderson model is regained.

\section{\label{sec:con} Conclusion}

\indent In this paper we examined the effect of applied uniaxial strain on the ferromagnetic exchange instability due to long- 
range Coulomb interaction between the Dirac fermions and also on the formation of localized magnetic moment on an adatom (impurity) 
in graphene. Our work was motivated by a broad aim to understand the behavior of itinerant as well as localized magnetism in 
deformed graphene. \\
\indent We derived the electronic dispersion of the anisotropic Dirac fermions and considered that the electrons in graphene are 
interacting via long- range Coulomb interaction. We calculated the total energy difference between the ferromagnetic (polarized) 
and paramagnetic (unpolarized) phases within the Hartree-Fock approximation and determined the necessary condition for the 
occurence of the exchange instability or the paramagnetic to ferromagnetic transition as a function of anisotropy. In the 
undeformed case, Peres \textit{et al.} conjectured that large critical coupling constant, i.e. strength of interaction can lead 
to first order magnetic (Bloch) transition. We found that the value of critical coupling constant, which caused exchange 
instability, decreased with increasing applied uniaxial strain. For large deformations
our work indicates that  graphene can have a ferromagnetic 
ground state for relatively weak effective interaction ($\alpha 
\lesssim 2$) which is comparable to the case of suspended graphene \cite{castroneto1}. The effect of uniaxial strain on the electronic properties has already been studied 
experimentally but the magnetic properties of deformed monolayer graphene have not been explored. It would be very interesting 
to confirm our result experimentally. \\
\indent We also formulated the conditions for the existence of localized magnetic moments on an adatom in a deformed graphene. 
Using the single impurity Anderson model and within mean- field self- consistent approach, we systematically studied the polarization 
of the electrons in the localized level as a function of model parameters like hybridization strength ($V$), on- site Coulomb 
repulsion ($U$), localized energy level ($E_0$) and the applied strain ($\varepsilon$). We obtained magnetic/non-magnetic phase 
boundary for different values of scaled model parameters and found that for moderate hybridization the magnetic region shrunk 
drastically with increasing strain as compared to weak hybridization. This can be understood by the fact that the anisotropic 
cut- off energy kept on decreasing with increase in deformation, which resulted in an increase in effective hybridization ($\Delta$). 
This in turn increased the density of states at the localized level and the electrons at the impurity level became more de-localized. 
The features predicted in this work can find possible applications in the field of carbon-based spintronics.

\section{\label{sec:ack} Acknowledgments} 

We thank Vitor M. Pereira for invaluable discussions and suggestions. This work was supported by DOE grant DE-FG02-08ER46512. AHCN acknowledges the NRF-CRP award "Novel 2D materials with tailored properties: beyond graphene" (R-144-000-295-281).

\end{document}